\documentclass[10pt,a4paper,twoside]{article}

\usepackage{amssymb,amsmath,multirow,epstopdf,graphicx}
\usepackage{amsthm}
\usepackage{graphics}
\usepackage{color}
\newtheorem{theorem}{Theorem}[section]

\newtheorem{lemma}[theorem]{Lemma}

\newtheorem{charact}[theorem]{Characterization}

\begin{document}
\title{New $L^2$-type exponentiality tests}

\author{Marija Cupari\'c\thanks{marijar@matf.bg.ac.rs},  Bojana Milo\v sevi\'c\footnote{bojana@matf.bg.ac.rs} , Marko Obradovi\' c \footnote{marcone@matf.bg.ac.rs}
         \\\medskip
{\small Faculty of Mathematics, University of Belgrade, Studenski trg 16, Belgrade, Serbia}}

\date{}

\maketitle
\begin{abstract}
    We introduce new  consistent and scale-free goodness-of-fit tests for the exponential distribution based on Puri-Rubin  characterization. For the construction of test statistics we employ  weighted $L^2$ distance between $V$-empirical Laplace transforms of random variables that appear in the characterization. The resulting test statistics are
	  degenerate V-statistics with estimated parameters.
	   	   	  We compare our tests, in terms of the Bahadur efficiency, to the likelihood ratio test, as well as some recent characterization based goodness-of-fit tests for the exponential distribution. We also compare the powers of our tests to the powers of some recent and classical exponentiality tests. In both criteria, our tests are shown to be strong and outperform most of their competitors.
\end{abstract}

{\small \textbf{ keywords:}  goodness-of-fit; exponential distribution; Laplace transform; Bahadur efficiency; V-statistics 

\textbf{MSC(2010):} 62G10, 62G20}

\renewcommand{\baselinestretch}{1.2}
\bigskip

\section{Introduction}
\label{intro}

The exponential distribution is one of most widely studied distributions in theoretical and applied statistics. Many models assume exponentiality of the data. Ensuring that those models can be used is of a great importance. 
 For this reason, a great variety of goodness of fit tests for the particular case of the exponential distribution, have been proposed in literature.


Different construvtions have been used to build test statistics. They are mainly based on empirical counterparts of some special properties of the exponential distribution.   Some  of those tests employ properties connected to different  integral transforms such as: characteristic functions (see e.g. \cite{henze1992new}, \cite{henze2002goodness}, \cite{henze2005}); Laplace transforms (see e.g. \cite{henze2002tests}, \cite{klar2003test}, \cite{meintanis2007testing}); and other integral transforms (see e.g. \cite {klar2005tests}, \cite{meintanis2008tests}). Other properties include maximal correlations (see \cite{grane2009location}, \cite{grane2011directional}), entropy  (see \cite{alizadeh2011testing}), etc.
 
 The simple form of the exponential distribution gave rise to many equidistribution type characterizations. The equality in distribution can be expressed in many ways (equality of  distribution functions, densities, integral transforms, etc.). This makes them  suitable for building different types of test statistics. Such tests have become very popular in recent times, as they are proven to be rather efficient. Tests that use U-empirical and V-empirical distribution functions, of integral-type (integrated difference) and supremum-type, can be found in \cite{NikVol}, \cite{volkova2015goodness}, \cite{jovanovic}, \cite{Publ}, \cite{bojanaMetrika}, \cite{nikitin2016efficiency}. A class of weighted integral-type tests that uses U-empirical Laplace transforms is presented in \cite{MilosevicObradovicPapers}. 
 
Motivated by the power and efficiency of those tests, here we
create a similar test based on an equidistribution
characterization.  The test statistics measure the distance between two V-empirical Laplace transforms of the random variables that appear in the characterization, but, for the first time, using  weighted $L^2$-distance. This guarantees the  consistency of the test against all alternatives.

The paper is organized as follows. In Section \ref{sec: sec2} we introduce the test statistics and derive their asymptotic properties. In Section \ref{sec: sec3} we calculate the approximate Bahadur slope of our tests, for different close alternatives, and inspect the impact of the tuning parameter to the efficiencies of the test. We also compare the proposed tests to their recent competitors, via approximate local relative  Bahadur efficiency. In Section \ref{sec: sec5} we conduct a power study. 
We obtain empirical powers of our tests, against different common alternatives, and compare them to some recent and classical exponentiality tests. We also apply an algorithm for data driven selection of tuning parameter and obtain the corresponding powers in small sample case.
\section{Test statistic}\label{sec: sec2}
Puri and Rubin \cite{puri1970characterization} proved the following characterization theorem.

\begin{charact}\label{puri}
	Let $X_1$ and $X_2$ be two independent copies of a random variable $X$ with pdf $f(x)$. Then $X$ and $|X_1-X_2|$ have the same distribution, if and only if  for some $\lambda>0$, $f(x)=\lambda e^{-\lambda x}$, for $x\geq 0$.
\end{charact}

Let $X_1,X_2,...,X_n$ be independent copies of a non-negative random variable $X$ with unknown distribution function $F$. We consider the transformed sample $Y_i=\hat{\lambda}X_i,\;i=1,2..,n.$, where $\hat{\lambda}$ is the reciprocal sample mean. For testing the null hypothesis $H_0: \;F(x)=1-e^{-\lambda x},\; \lambda>0,$
in  view of the characterization \ref{puri}, we propose the following family of test statistics, depending on the tuning parameter $a>0$,
\begin{equation}\label{testM}
M_{n,a}(\hat{\lambda})=\int_0^{\infty}\left(L^{(1)}_n(t)-L^{(2)}_n(t)\right)^2e^{-at}dt, 
\end{equation}
where 
\begin{align*}
L_n^{(1)}(t)&=\frac{1}{n}\sum_{i_1=1}^ne^{-tY_{i_1}}\\
L_n^{(2)}(t)&=\frac{1}{n^{2}}\sum_{i_1,i_2=1}^ne^{-t |Y_{i_1}-Y_{i_2}|}
\end{align*}
 are V-empirical Laplace transforms of $Y_1$ and $|Y_1-Y_2|$ respectively.
 
In order to explore the asymptotic properties  we rewrite \eqref{testM} as 
 \begin{equation*}
 \begin{aligned}
 M_{n,a}(\hat{\lambda})&=\int_0^{\infty}\left(\frac{1}{n^{2}}\sum_{i_1=1}^ne^{-tX_{i_1}\hat{\lambda}}-\frac{1}{n^{2}}\sum_{i_1,i_2=1}^ne^{-t |X_{i_1}-X_{i_2}|\hat{\lambda}}\right)^2e^{-at}dt\\
 &=\frac{1}{n^{4}}\int_0^\infty\sum_{i_1,i_2,i_3,i_{4}}\left(e^{-tX_{i_1}\hat{\lambda}}-e^{-t |X_{i_1}-X_{i_2}|\hat{\lambda}}\right)\left(e^{-tX_{i_3}\hat{\lambda}}-e^{-t |X_{i_3}-X_{i_4}|\hat{\lambda}}\right)e^{-at}dt\\
 &=\frac{1}{n^{4}}\sum_{i_1,i_2,i_3,i_{4}}\int_0^\infty g(X_{i_1},X_{i_2},t;\hat{\lambda})g(X_{i_{3}},X_{i_{4}},t;\hat{\lambda})e^{-at}dt\\
 &=\frac{1}{n^{4}}\sum_{i_1,i_2,i_3,i_{4}}h(X_{i_1},X_{i_2},X_{i_3},X_{i_{4}},a;\hat{\lambda}),
 \end{aligned}
 \end{equation*}
 where $\hat{\lambda}=\bar{X}^{-1}$ is a consistent estimator of $\lambda.$
 
\begin{figure}[ht]
	\begin{center}
\includegraphics[scale=0.5]{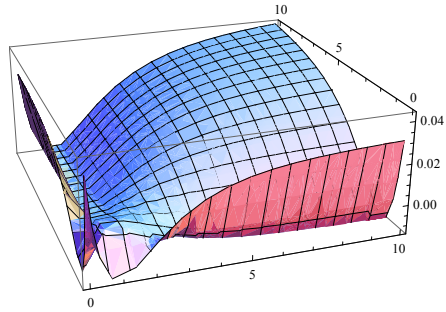}
	\caption{Second projection $\tilde{h}_2(x,y,1)$}\label{fig:projMarija}
 \end{center}
  \end{figure}

 Let's focus, for a moment, on $M_{n,a}(\lambda)$, for a fixed $\lambda>0$. Notice that  $M_{n,a}(\lambda)$ is a $V$-statistic with kernel $h$. Moreover, under the null hypothesis its distribution does not depend on $\lambda$, so we may assume $\lambda=1$.
 It is easy to show that its first projection on a basic observation is equal to zero. After some calculations, one can obtain its second projection given by
 \begin{align*}
    \tilde{h}_2(x,y,a)&=
    E(h(X_1,X_2,X_3,X_4,a|X_1=x,X_2=y)\\&=
    -\frac{1}{2}+\frac{1}{3}(e^{-x}+e^{-y})+\frac{1}{6}e^{a-x-y}\text{Ei}(-a) \Big(a (e^x-2) (e^y-2)-e^x-e^y+4\Big)\\&+\frac{1}{6}e^{-a-x-y}\Big(\text{Ei}(a) (4 a+e^x+e^y-4) - (\text{Ei}(a+x) (4 (a+x-1)+e^y)\\&+\text{Ei}(a+y) (4 (a+y-1)+e^x)-4 (a+x+y-1) \text{Ei}(a+x+y))\Big)+\frac{1}{6(a+x+y)},
\end{align*}
where $\text{Ei}(x)=-\int_{-x}^\infty \frac{e^{-t}}{t}dt$ is the exponential integral. The function $\tilde{h}_2$ is  non-constant for any $a>0$. Its plot, for $a=1$, is shown in Figure \ref{fig:projMarija}. Hence, the kernel $h$ is degenerate with degree 2. 

 The asymptotic distribution of $M_{n,a}(\hat{\lambda})$ is given in the following theorem.
 
 \begin{theorem}\label{raspodela}
 	Let $X_1,...,X_n$ be i.i.d. sample with distribution function $F(x)=1-e^{\lambda x}$ for some $\lambda>0$. Then
 	\begin{equation}\label{raspodelaT}
 	nM_{n,a}(\hat{\lambda})\overset{d}{\rightarrow}6\sum_{k=1}^\infty\delta_kW^2_k,
 	\end{equation}
 	where $\{\delta_k\}$  are the eigenvalues of the integral operator $\mathcal{M}_a$ defined by  
 	\begin{equation} \label{operatorA}
 	\mathcal{M}_{a}q(x)=\int_{0}^{+\infty}\tilde{h}_2(x,y,a)q(y)dF(y)
 	\end{equation}
 	  and  $\{W_{k}\}$ is the sequence  of i.i.d  standard Gaussian random variables.
 \end{theorem}
 
\begin{proof}
 Since the kernel $h$ is bounded and  degenerate,
 from the theorem for the asymptotic distribution of U-statistics with
degenerate kernels \cite[Corollary 4.4.2]{korolyuk}, and the Hoeffding representation of $V$-statistics, we get that, $M_{n,a}(1)$, being a $V$-statistic of degree 2, has the asymptotic distribution from \eqref{raspodelaT}. Hence, it suffices to show that $M_{n,a}(\hat{\lambda})$ and $M_{n,a}(1)$ have the same distribution.

 Our statistic $M_{n,a}(\hat{\lambda})$ can be  rewritten as
\begin{equation*}
\begin{aligned}
M_{n,a}(\hat{\lambda})&=\int_0^{\infty}\left(\frac{1}{n^{2}}\sum_{i_1,i_2=1}^ng(X_{i_1},X_{i_2},t,a;\hat{\lambda})\right)^2e^{-at}dt\\
&=\int_0^{\infty}V_n(\hat{\lambda})^2e^{-at}dt.
\end{aligned}
\end{equation*}
Here $V_n(\hat{\lambda})$  is a $V$-statistic of order 2 with estimated parameter, and kernel  $g(X_{i_1},X_{i_2},t,a;\hat{\lambda})$.

Since the function $g(x_{1},x_{2},t,a;\gamma)$ is continuously differentiable with respect to $\gamma$ at the point  $\gamma=\lambda$, the mean-value theorem gives us  
\begin{equation*}
V_n(\hat{\lambda})=V_n(\lambda)+(\hat{\lambda}-\lambda)\frac{\partial V_n(\gamma)}{\partial\gamma}|_{\gamma=\lambda^*},
\end{equation*}
for some $\lambda^*$ is between $\lambda$ and  $\hat{\lambda}$.

Using the Law of large numbers for V-statistics \cite[6.4.2.]{Serfling}, we have that
$\frac{\partial V_n(\gamma)}{\partial\gamma}$ converges to

\begin{equation*}E\left(t|X_1-X_2|e^{-t|X_{1}-X_{2}|\gamma}-tX_{1}e^{-tX_{1}\gamma}\right)=0.
\end{equation*}
Since $\sqrt{n}(\hat{\lambda}-\lambda)$ is stochastically bounded,
we conclude that statistics $\sqrt{n}V_n(\hat{\lambda})$ and $\sqrt{n}V_n(1)$ are  asymptotically equally distributed.  Therefore, $nM_{n,a}(\hat{\lambda})$ and $nM_{n,a}(1)$ will have the same limiting distribution, which completes the proof.
\end{proof}

\section{Local Approximate Bahadur efficiency}\label{sec: sec3}
One way to compare tests is to calculate their relative Bahadur efficiency.  We briefly present it here. For more details  we refer to \cite{bahadur1971} and \cite{nikitin}.

For two tests
with the same null and alternative hypotheses, $H_0: \theta\in \Theta_0$ and $H_1: \theta \in \Theta_1$, the
asymptotic relative Bahadur efficiency is defined as the ratio of sample sizes needed
to reach the same test power, when the level of significance approaches zero. For two sequences of test statistics, it can be
expressed as the ratio of Bahadur exact slopes, functions proportional to exponential
rates of decrease of their sizes, for the increasing number of observations and a fixed alternative. The calculation of these slopes depends on large
deviation functions which are often hard to obtain.
For this reason, in many situations, the tests are compared using the approximate Bahadur
efficiency, which is shown to be a good approximation in the local case (when $\theta\to \partial\Theta_0$).

Suppose that $T_n=T_n(X_1,..., X_n)$ is a test statistic with its large values being
significant. Let the limiting distribution function of  $T_n$, under $H_0$, be $F_T$, whose tail behavior is given by  $\log(1-F_T (t))=-\frac{a_T t^2}{2}(1 + o(1))$, where $a_T$ is positive
real number, and $o(1) \to 0$ as $t\to \infty$. 
Suppose also that the limit in probability
$\lim_{n\to \infty} T_n/
\sqrt{n} = b_T(\theta)>0$ exists for $\theta \in \Theta_1.$ Then the relative approximate Bahadur efficiency of $T_n$, with respect to another test
statistic $V_n$ (whose large values are significant), is
\begin{equation*}
    e^*_{V,T}(\theta)=\frac{c^*_{V}(\theta)}{c^*_{T}(\theta)},
\end{equation*}
where $c^*_{T}(\theta)=a_Tb_T^2(\theta)$ i $c^*_{V}(\theta)=a_Vb_V^2(\theta)$ are approximate Bahadur slopes of $T_n$ and $V_n$, respectively. 

We may suppose, without loss of generality, that $\Theta_0=\{0\}$. Consequently, the approximate local relative  Bahadur efficiency is  given by
\begin{equation*}
    e^*_{V,T}=\lim_{\theta \to 0}e^*_{V,T}(\theta).
\end{equation*}

Let $\mathcal{G}=\{G(x,\theta),\;\theta>0\}$ be a family of alternative distribution functions such that $G(x,\theta)=1-e^{-\lambda x}$, for some $\lambda>0$, if and only if $\theta=0$, and the regularity conditions for V-statistics with weakly degenerate kernels from \cite[Assumptions WD]{nikitinMetron} are satisfied.

The logarithmic tail behaviour of the limiting distribution of $M_{n,a}(\hat{\lambda})$, under the null hypothesis, is derived in the following lemma.

\begin{lemma}
For the statistic $M_{n,a}(\hat{\lambda})$ and the given alternative density $g(x,\theta)$ from $\mathcal{G},$ the
Bahadur approximate slope satisfies the relation
$c_M(\theta)\sim\frac{b_M(\theta)}{6\delta_1}$, where $b_M(\theta)$ is the limit in $P_{\theta}$ probability of $M_{n,a}(\hat{\lambda})$, and $\delta_1$ is the largest eigenvalue of the
sequence $\{\delta_k\}$ from \ref{raspodela}.
\end{lemma}
\begin{proof}
Using the result of Zolotarev \cite{Zolotarev}, we have that the logarithmic tail behavior
of limiting distribution  function of $\tilde{M}_{n,a}(\hat{\lambda})=\sqrt{nM_{n,a}(\hat{\lambda})}$ is

\begin{equation*}
    \log(1-F_{\tilde{M}_a}(t))=-\frac{t^2}{12\delta_1}+o(t^2),\;\;t\to \infty.
\end{equation*}
Therefore, we obtain that $a_{\tilde{M}_a}=\frac{1}{6\delta_1}.$
The limit in probability $P_{\theta}$ of $\tilde{M}_{n,a}(\hat{\lambda})/\sqrt{n}$ is

\begin{equation*}
b_{\tilde{M}_{a}}=\sqrt{b_{M}(\theta)}.    
\end{equation*}
Inserting this into the expression for Bahadur slope, we complete the proof.
\end{proof}

The limit in probability of our test statistic, under a close alternative, can be derived using the following Lemma.

\begin{lemma}
For a given alternative density $g(x;\theta)$ whose distribution belongs to $\mathcal{G}$, we have that the limit in probability of the statistic $M_{n,a}(\hat{\lambda})$ is 	

\begin{equation*}
b_M(\theta)=6\int\limits_{0}^{\infty}\int\limits_{0}^{\infty}\tilde{h}_2(x,y)f(x)f(y)dxdy\cdot\theta^2+o(\theta^2), \theta\rightarrow0,
\end{equation*}
where $f(x)=\frac{\partial}{\partial\theta}g(x;\theta)|_{\theta=0}.$
\end{lemma}
\begin{proof}
For brevity, let us denote  $\boldsymbol{x}=(x_1,x_2,x_3,x_{4})$ and $\boldsymbol{G}(\boldsymbol{x};\theta)=\prod_{i=1}^{4}G(x_i;\theta)$.
Since $\bar{X}$ converges almost surely to its
expected value $\mu(\theta)$, using the Law of large numbers for $V$-statistics with estimated parameters (see \cite{iverson}), we  have  that $M_{n,a}(\hat{\lambda})$ converges to
\begin{equation*}
\begin{aligned}
	b_M(\theta)&=E_{\theta}(h(\boldsymbol{X},a;\mu(\theta)))\\&=\int_{{(R^+)}^{4}}\Big(\frac{\mu(\theta)}{x_1+x_{3}+a\mu(\theta)}-\frac{\mu(\theta)}{x_{3}+|x_{1}-x_{2}|+a\mu(\theta)}\\&-\frac{\mu(\theta)}{x_{1}+|x_{3}-x_{4}|+a\mu(\theta)}+\frac{\mu(\theta)}{|x_{1}-x_{2}|+|x_{3}-x_{4}|+a\mu(\theta)}\Big)d\boldsymbol{G}(\boldsymbol{x};\theta).
\end{aligned}
\end{equation*}
We may assume that $\mu(0)=1$ due to the scale freeness of test statistic under the null hypothesis. After some calculations we get that $b'_M(0)=0.$ Next, we obtain that
\begin{equation*}
\begin{split}
b''(0)&
=\int_{(R^+)^{4}}h(\boldsymbol{x},a;1)\frac{\partial^2}{\partial\theta^2}d\boldsymbol{G}(\boldsymbol{x},0)
=6\int_{(R^+)^2}\tilde{h}_2(x,y)f(x)f(y)dxdy.
\end{split}
\end{equation*}

Expanding $b_M(\theta)$ into the Maclaurin series we complete the proof. 
\end{proof}
To calculate the efficiency one needs to find $\delta_1$, the largest eigenvalue. Since we can not obtain it analytically, we use the following approximation, introduced in \cite{bozin}.

It can be shown that $\delta_1$ is the limit od the sequence of the largest eigenvalues of
linear operators defined by $(m+1)\times(m+1)$ matrices $M^{(m)}=||m_{i,j}^{(m)}||,\; 0\leq i\leq m, 0\leq j\leq m$, where
\begin{equation}\label{MatAppr}
m_{i,j}^{(m)}=\tilde{h}_2\bigg(\frac{Bi}{m},\frac{Bj}{m}\bigg)\sqrt{e^{\frac{B(i)}{m}}-e^{\frac{B(i+1)}{m}}}\cdot\sqrt{e^{\frac{B(j)}{m}}-e^{\frac{B(j+1)}{m}}}\cdot \frac{1}{1-e^{-B}},
\end{equation}
when $m$ tends to infinity   and $F(B)$ approaches 1.

In Table \ref{fig:EigenApprox}, we present the largest eigenvalues for $a=$0.5, 1, 2 and 5, obtained using \eqref{MatAppr} with $m=4500$ and $B=10$.

\begin{table}[!hbt]
    \centering
    \caption{Approximate eigenvalues of $\mathcal{M}_a$}
    \label{fig:EigenApprox}
    \begin{tabular}{c|cccc}
        $a$ &  0.5 & 1 & 2 & 5\\
        $\delta_1$ & $1.32\cdot10^{-2}$ & $5.32\cdot10^{-3}$ & $1.73\cdot10^{-3}$ & $2.80\cdot10^{-4}$  
    \end{tabular}
\end{table}

\subsection{Efficiencies with respect to LRT}

Lacking a theoretical upper bound, the approximate Bahadur slopes are often compared
(see e.g. \cite{meintanis2007testing}) to the approximate Bahadur slopes of the
likelihood ratio tests (LRT), which are known to be optimal parametric tests in terms of
Bahadur efficiency. Hence, we may consider the approximate relative  Bahadur efficiencies
against the LRT as a sort of "absolute" local approximate Bahadur efficiencies. We calculate it for the following alternatives:
\begin{itemize}
    \item a Weibull distribution with  density
    \begin{equation*}
        g(x,\theta)=e^{-x^{1+\theta}}(1+\theta)x^\theta,\theta>0,x\geq0;
    \end{equation*}
    \item a Gamma distribution with density
    \begin{equation*}
        g(x,\theta)=\frac{x^\theta e^{-x}}{\Gamma(\theta+1)},\theta>0,x\geq0;
    \end{equation*}
    \item a Linear failure rate distribution with density 
    \begin{equation*}
        g(x,\theta)=e^{-x-\theta\frac{x^2}{2}}(1+\theta x),\theta>0,x\geq0;
    \end{equation*}
    \item a mixture of exponential distributions with negative weights (EMNW($\beta$)) with density (see \cite{jevremovic1991note})
    \begin{equation*}
        g(x,\theta)=(1+\theta)e^{-x}-\theta\beta e^{-\beta x},\theta\in\left(0,\frac{1}{\beta-1}\right],x\geq0;
    \end{equation*}
\end{itemize}

It is easy to show that all densities given above belong to family $\mathcal{G}.$

\begin{figure}[!h]
	\begin{center}
	\includegraphics[scale=0.5]{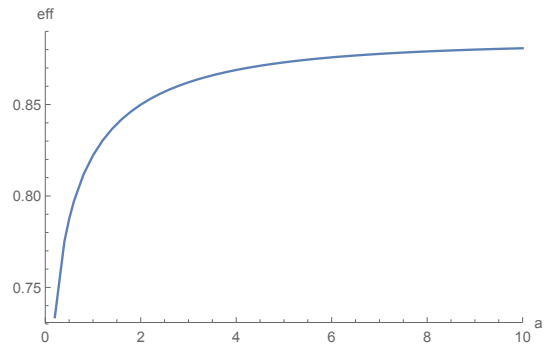}
		\caption{Local approximate Bahadur efficiencies w.r.t. LRT for a Weibull alternative}
		\label{fig: appslopeW}
	\end{center}
\end{figure}

\begin{figure}[!h]
	\begin{center}
		\includegraphics[scale=0.5]{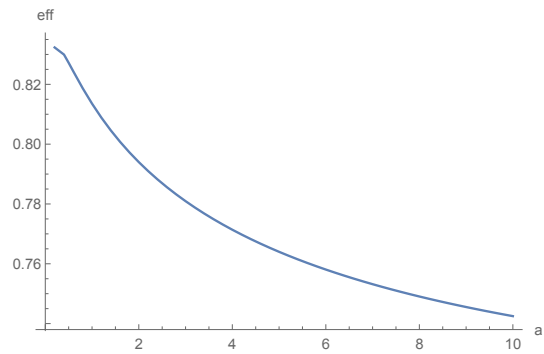}
	\caption{Local approximate Bahadur efficiencies w.r.t. LRT for a gamma alternative}
	\label{fig: appslopeG}
	\end{center}
\end{figure}

\begin{figure}[!h]
	\begin{center}
	\includegraphics[scale=0.5]{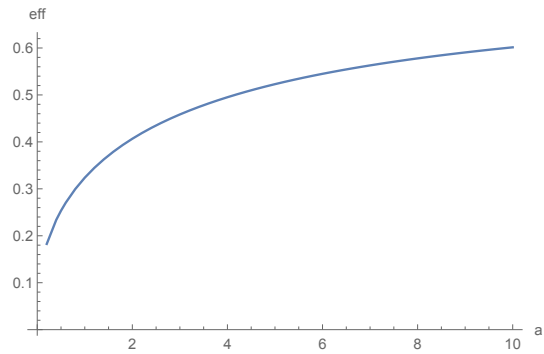}
		\caption{Local approximate Bahadur efficiencies w.r.t. LRT for a linear failure rate alternative}
		\label{fig: appslopeL}
	\end{center}
\end{figure}

\begin{figure}[!h]
	\begin{center}
	\includegraphics[scale=0.5]{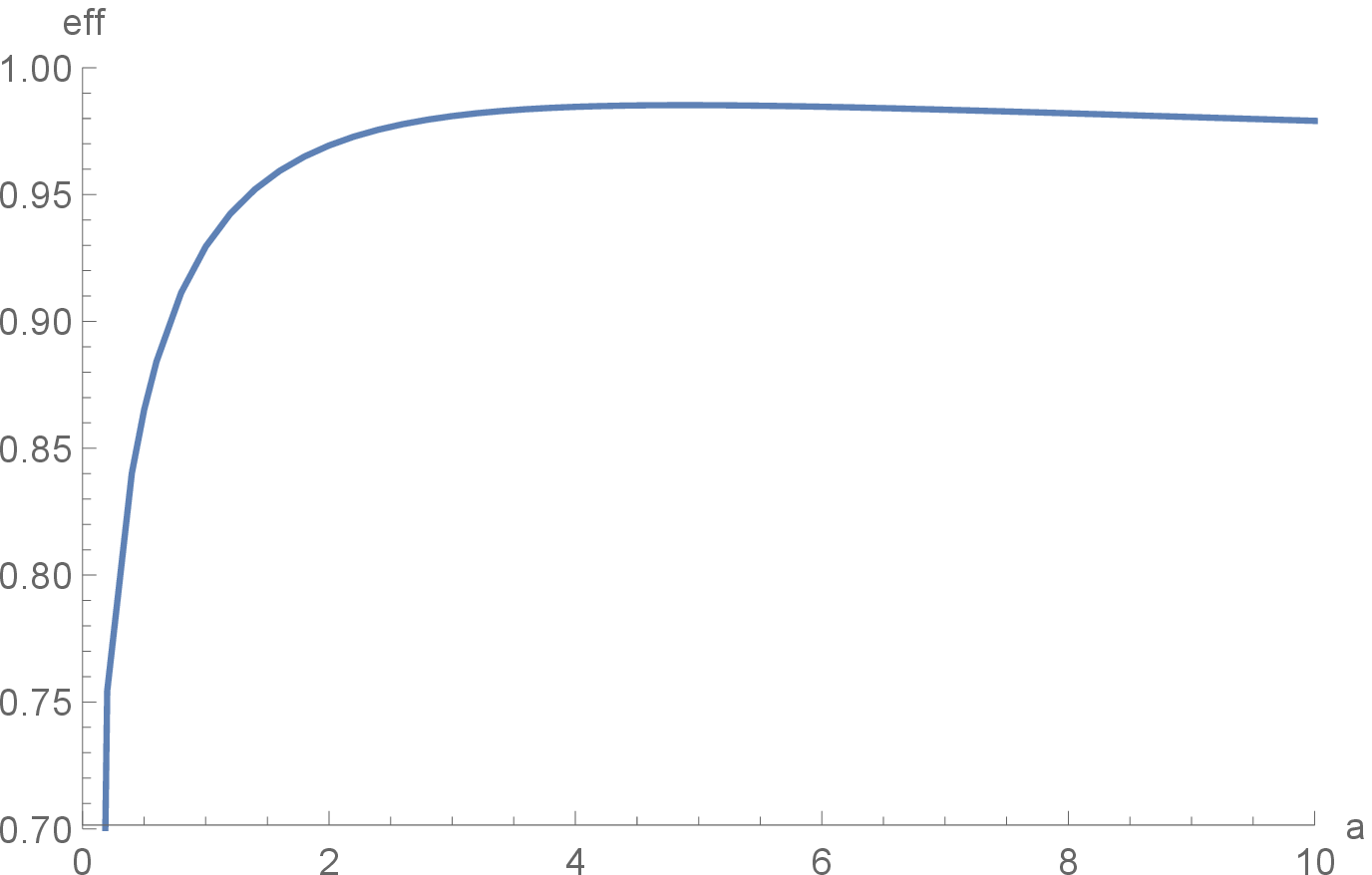}
		\caption{Local approximate Bahadur efficiencies w.r.t. LRT for an EMNW(3) alternative}
		\label{fig: appslopeM}
	\end{center}
\end{figure}

The efficiencies, as functions of the tuning parameter $a$, are shown on Figures \ref{fig: appslopeW}-\ref{fig: appslopeM}.

We can notice that the local efficiencies range from reasonable to high, and for some values of $a$ they are very high. Also, their behaviour with respect to the tuning parameter $a$ is very different. In the cases of Weibull and Linear failure rate  alternatives (Figures \ref{fig: appslopeW} and \ref{fig: appslopeL}), they are increasing functions of $a$, while in the Gamma case (Figure \ref{fig: appslopeG}), the function is decreasing. In the case of EMNW(3) (Figure \ref{fig: appslopeM}), the efficiencies increase up to a certain point and then decrease. 

\subsection{Comparison of efficiencies}

In this section, we calculate the local approximate Bahadur relative efficiency of our tests against some recent, characterization based integral-type tests, for the previously mentioned alternatives.

The  characterizations are of the equidistribution type and take the following form.

\emph{Let $X_1,...,X_{\max(m,p)}$ be i.i.d  with d.f. $F$,  $\omega_1: R^m\mapsto R^1$ and $\omega_2:R^p\mapsto R^1$ two sample functions.
	Then the following relation holds
	$$\omega_1(X_1,...,X_m)\overset{d}{=}\omega_2(X_1,...,X_p)$$
	if and only if $F(x)=1-e^{-\lambda x}$, for some $\lambda>0.$}

Notice that the Puri-Rubin characterization  is an example of such characterizations.

The first  class of competitor tests consists of the integral-type tests with test statistic 
\begin{equation*}
    I_n=\int_{0}^{\infty}\Big(G_n^{(1)}(t)-G_n^{(2)}(t)\Big)dF_n(t),
\end{equation*}
where $G_n^{(1)}(t)$ and $G_n^{(2)}(t)$ are $V$-empirical distribution functions of $\omega_1$ and $\omega_2,$ respectively.

In particular, we consider the following integral-type test statistics:
\begin{itemize}
    \item $I_{n,k}^{(1)}$, proposed in \cite{jovanovic}, based on the Arnold and Villasenor characterization, where $\omega_1(X_1,...,X_k)=\max(X_1,...,X_k)$ and $\omega_2(X_1,...,X_k)=X_1+\frac{X_2}{2}+\cdots \frac{X_k}{k}$ (see \cite{arnold2013exponential}, \cite{milovsevic2016some});
    \item $I_{n}^{(2)}$, proposed in \cite{Publ}, based on the Milo\v sevi\'c-Obradovi\'c characterization,  where $\omega_1(X_1,X_2)=\max(X_1.X_2)$ and $\omega_2(X_1,X_2,X_3)=\min(X_1,X_2)+X_3$ (see \cite{milovsevic2016some});
    \item $I_n^{(3)}$, proposed in \cite{bojanaMetrika}, based on the Obradovi\'c characterization, where $\omega_1(X_1,X_2,X_3)=\max(X_1,X_2,X_3)$ and $\omega_2(X_1,X_2,X_3,X_4)=X_1+{\rm med}(X_2,X_3,X_4)$ (see \cite{obradovic2014three});
    \item $I_n^{(4)}$, proposed in \cite{volkova2015goodness}, based on the Yanev-Chakraborty characterization, where $\omega_1(X_1,X_2,X_3)=\max(X_1,X_2,X_3)$ and $\omega_2(X_1,X_2,X_3)=\frac{X_1}{3}+\max(X_2,X_3)$ (see \cite{yanev2013characterizations}).
\end{itemize}

We also consider integral-type tests of the form
\begin{equation*}
    J_{n,a}=\int_{0}^{\infty}\Big(L_n^{(1)}(t)-L_n^{(2)}(t)\Big)\bar{X}e^{-at}dt,
\end{equation*}
where $L_n^{(1)}(t)$ and $L_n^{(2)}(t)$ are $V$-empirical Laplace transforms of $\omega_1$ and $\omega_2$, respectively.
This approach has been originally proposed in \cite{MilosevicObradovicPapers}. There,  particular cases of Desu characterization, with $\omega_1(X_1)=X_1$ and  $\omega_2=2\min(X_1,X_2)$, and Puri-Rubin characterization were examined. We denote the corresponding tests statistics with $J_{n,a}^{\mathcal{D}}$ and $J_{n,a}^{\mathcal{P}}$, respectively. The results are presented in Table \ref{fig: efikasnosti}. We can notice that in most cases tests that employ $V$-empirical Laplace transforms are more efficient than those based on $V$-empirical distribution functions. On the other hand, new tests are comparable with $J_{n,a}^{\mathcal{P}}$ and more efficient than  $J_{n,a}^{\mathcal{D}}$. 

\begin{table}[!htb]
	\centering
	\caption{Relative Bahadur efficiency of $M_{n,a}$ with respect to its competitors}
\label{fig: efikasnosti}
\begin{tabular}{cccccc}
    & $a$ & 0.5 & 1 & 2 & 5 \\\hline
   {$I_{n,2}^{(1)}$} & $Weibull$ & 1.27 & 1.33 & 1.37 & 1.42 \\
    & $Gamma$ & 1.14 & 1.13 & 1.10 & 1.06\\
    & $LFR$ & 2.44 & 3.13 & 3.93 & 5.08\\
    & $EMNW(3)$ & 1.25 & 1.34 & 1.40 & 1.42 \\ \hline
   {$I_{n,3}^{(1)}$} & $Weibull$ & 1.19 & 1.24 & 1.28 & 1.32 \\
    & $Gamma$ & 1.17 & 1.15 & 1.12 & 1.09\\
    & $LFR$ & 1.59 & 2.04 & 2.56 & 3.31\\
    & $EMNW(3)$ & 1.08 & 1.17 & 1.22 & 1.23\\ \hline
     {$I_{n}^{(2)}$} & $Weibull$ & 1.05 & 1.10 & 1.14 & 1.17 \\
    & $Gamma$ & 1.04 & 1.02 & 1.00 & {0.97}\\
    & $LFR$ & 1.22 & 1.56 & 1.96 & 2.53\\
    & $EMNW(3)$ & 1.02 & 1.10 & 1.15 & 1.17 \\ \hline
   {$I_{n}^{(3)}$} & $Weibull$ & 1.06 & 1.10 & 1.14 & 1.18 \\
    & $Gamma$ &1.18 & 1.16 & 1.14 & 1.10\\
    & $LFR$ & 0.82 & 1.05 & 1.32 & 1.71\\
    & $EMNW(3)$ & {0.94} & 1.02 & 1.06 & 1.08\\ \hline
   {$I_{n}^{(4)}$} & $Weibull$ & 1.21 & 1.27 & 1.31 & 1.35\\
    & $Gamma$ &1.30 & 1.28 & 1.25 & 1.21\\
    & $LFR$ & 1.23 & 1.57 & 1.98 & 2.56\\
    & $EMNW(3)$ & 1.04 & 1.12 & 1.16 & 1.18\\ \hline
    {$J^{\mathcal{P}}_{n,a}$} & $Weibull$ &{0.97} & 0.97 & 1.01 & 1.00\\
    & $Gamma$ &{0.98} & {0.99} & 1.00 & 1.02\\
    & $LFR$ & 0.97 & 0.93 & 0.91 & {0.93}\\
    & $EMNW(3)$ &0.97 & {0.98} & {0.99} & 1.00\\ \hline
    {$J_{n,a}^{\mathcal{D}}$} & $Weibull$ &1.00 & {0.95} & {0.93} & {0.95}\\
    & $Gamma$ &2.16 & 1.64 & 1.33 & 1.13\\
    & $LFR$ & 1.17 & 1.07 & 1.01 & 0.99\\
    & $EMNW(3)$ &1.42 & 1.18  & 1.06 & {0.99}\\ \hline
\end{tabular}
\end{table}

\section{ Power study }\label{sec: sec5}
In this section we compare the empirical powers of our tests with those of some common competitors, listed in \cite{henze2005} and \cite{MilosevicObradovicPapers}. The Monte Carlo study is done for small sample size $n=20$ and the moderate sample size $n=50$, with $N=10000$ replicates, for level of significance $\alpha=0.05.$ 

The powers are presented in Tables \ref{fig: comparison20} and \ref{fig: comparison50}. The labels used are identical to the ones in \cite{henze2005} and \cite{MilosevicObradovicPapers}.
\begin{table}[htbp]
	\centering
	\caption{Percentage of rejected hypotheses for  $n=20$ }
	\resizebox{\textwidth}{!}{
		\begin{tabular}{ccccccccccc}
			Alt. & $W(1.4)$ & $\Gamma(2)$ & $HN$ & $U$ & $CH(0.5)$ & $CH(1)$ & $CH(1.5)$ & $LF(2)$ & $LF(4)$ &  $EW(1.5)$\\\hline
			$EP$              & 36 & 48 & 21 & 66 & 63 & 15 & 84 & 28 & 42  & 45\\
			$\overline{KS}$ & 35 & 46 & 24 & 72 & 47 & 18 & 79 & 32 & 44 &  48\\
			$\overline{CM}$ & 35 & 47 & 22 & 70 & 61 & 16 & 83 & 30 & 43 &  47\\
			$\omega^2$      & 34 & 47 & 21 & 66 & 61 & 14 & 79 & 28 & 41 &  43\\
			$KS$              & 28 & 40 & 18 & 52 & 56 & 13 & 67 & 24 & 34 & 35\\
			$KL$              & 29 & 44 & 16 & 61 & 77 & 11 & 76 & 23 & 34 & 37\\
			$S$              & 35 & 46 & 21 & 70 & 63 & 15 & 84 & 29 & 42 & 46\\
			$CO$              & 37 & 54 & 19 & 50 & {\bf 80} & 13 & 81 & 25 & 37 & 37\\
			$J^{\mathcal{D}}_{n,1}$  & 42 & 64 & 20 & 45 & 15 & 15 & 15 & 29 & 40 & 36\\
			$J^{\mathcal{D}}_{n,2}$  & 47 & 66 & 25 & 59 & 18 & 19 & 18 & 33 & 48 & 46\\
			$J^{\mathcal{D}}_{n,5}$  & 48 & 64 & 28 & 70 & 20 & 21 & 21 & 36 & 52 & 53\\
			$J^{\mathcal{P}}_{n,1}$  & 49 & 65 & 29 & 73 & 21 & 22 & 21 & 38 & 51 & 54 \\
			$J^{\mathcal{P}}_{n,2}$  & {\bf 50} & 64 & 31 & 77 & 21 & 21 & 23 & 40 & 54 & 57 \\
			$J^{\mathcal{P}}_{n,5}$  & 48 & 62 & 32 & 79 & 23 & {\bf 23} & 23 & {\bf 41} & {\bf 56} & {\bf 58} \\
			$M_{n,0.5}$  & 46 & 66 & 25 & 64 & 19 & 18 & 19 & 35 & 49 & 46\\
			$M_{n,1}$  & 49 &	66 & 28 & 72 & 21 & 21 & 21	& 38 & 52 & 53 \\
			$M_{n,2}$  & {\bf 50} &	{\bf 67} & 31 & 75 & 22 & {\bf 23} & 23 & 40 & 55 & 56 \\
		    $M_{n,5}$  & 48 &	62 & {\bf 32} & {\bf 80} & 22 &{\bf 23} & 24	&  40 & {\bf 56} &	{\bf 58} \\
		  \end{tabular}
	}
	\label{fig: comparison20}
\end{table}
\begin{table}[htbp]
	\centering
	\caption{Percentage of rejected hypotheses for $n=50$}
	\resizebox{\textwidth}{!}{
		\begin{tabular}{ccccccccccc}
			Alt. & $W(1.4)$ & $\Gamma(2)$ & $HN$ & $U$ & $CH(0.5)$ & $CH(1)$ & $CH(1.5)$ & $LF(2)$ & $LF(4)$ & $EW(1.5)$\\\hline
			EP              & 80 & 91 & 54 & 98 & 94 & 38 & {\bf 100} & 69 & 87 &  90\\
			$\overline{KS}$ & 71 & 86 & 50 &{\bf  99} & 90 & 36 & {\bf 100} & 65 & 82 & 88\\
			$\overline{CM}$ & 77 & 90 & 53 & {\bf 99} & 94 & 37 & {\bf 100} & 69 & 87 & 90\\
			$\omega^2$      & 75 & 90 & 48 & 98 & 95 & 32 &{\bf  100} & 64 & 83 & 86\\
			KS              & 64 & 83 & 39 & 93 & 92 & 26 & 98 & 53 & 72 & 75\\
			KL              & 72 & 93 & 37 & 97 & {\bf 99} & 23 & {\bf 100} & 54 & 75 & 79\\
			S              & 79 & 90 & 54 & {\bf 99} & 94 & 38 & {\bf 100} & 69 & 87 & 90\\
			CO              & 82 & 96 & 45 & 91 & {\bf 99} & 30 & {\bf 100} & 60 & 80 &  78\\
			$J^{\mathcal{D}}_{n,1}$  & 78 & 96 & 36 & 76 & 23 & 24 & 23 & 51 & 71 &  64\\
			$J^{\mathcal{D}}_{n,2}$  & 83 & {\bf 97} & 46 & 90 & 31 & 30 & 31 & 62 & 83 & 79\\
			$J^{\mathcal{D}}_{n,5}$  & 86 & {\bf 97} & 55 & 97 & 41 & 40 & 40 & 72 & 89 & 89\\
			$J^{\mathcal{P}}_{n,1}$  & 85 & 96 & 54 & 97 & 38 & 38 & 38 & 70 & 87  & 87 \\
			$J^{\mathcal{P}}_{n,2}$  & 86 & 96 & 59 & 98 & 41 & 42 & 42 & 73 & 89 & 90 \\
			$J^{\mathcal{P}}_{n,5}$  & 86 & 96 & 63 & {\bf 99} & 46 & 46 & 45 & 77 & {\bf 91}  & {\bf 93} \\
			$M_{n,1}$  & 85 & {\bf 97} & 54 & 97 & 38 & 38 & 38 & 69 & 87 & 86 \\
			$M_{n,2}$  & 86 & 96 & 57 & 98 & 41 & 41 & 41 & 73 & 89 & 90 \\
			$M_{n,5}$  & {\bf 87} & 96 & {\bf 63} & {\bf 99} & 45 & 45 & 45 & {\bf 76} & {\bf 91} & {\bf 93} \\
         \end{tabular}
	}
	\label{fig: comparison50}
\end{table}

It can be noticed that in the majority of cases the tests based on $V$-empirical Laplace transforms are most powerful. Among them, those tests that are based on same characterization have more or less the same empirical powers, and the similar sensibility to the change of tuning parameter, for each considered alternative. 

\subsection{On a data-dependent choice of tuning parameter}

The powers of proposed tests depend on the values of tuning parameter $a$, and  the well-chosen value of $a$ would help us make the right decision. However, since the "right" value of $a$ is rather different for different alternatives, a general conclusion, which $a$ is most suitable in practice, can not be made. Hence, in what follows, we present an algorithm for data driven selection of tuning parameter, proposed initially by Allison and Santana \cite{allison}: 
\begin{enumerate}
    \item fix a grid of positive values of $a, (a_1,...,a_k)$;
    \item obtain a bootstrap sample $\boldsymbol{X}_n^*$ from empirical distribution function of $\boldsymbol{X}_n$;
    \item determine the value of test statistic $M_{n,a_i}, i=1,...,k,$ for the obtained sample; 
    \item repeat steps 2 and 3 $B$ times and obtain series of values of test statistics for every $a$, $M^*_{j,a_i}, i=1,...,k, j=1,...,B$;
    \item determine the empirical power of the test for every $a$, i.e. $$\hat{P}_{a_i}=\frac{1}{B}\sum_{j=1}^BI\{M_{j,a_i}\geq\check{C}_{n,a_i}(\alpha)\}, i=1,...,k;$$
    \item for the next calculation  $\hat{a}=\underset{a\in\{a_1,...,a_k\}}{argmax}\hat{P}_a$ will be used.
\end{enumerate}
   The critical value $\check{C}_{n,\hat{a}}$ is determined using the Monte Carlo procedure with $N_1$ replicates. Then, the empirical power of the test is determined based on the new sample from the alternative distribution
\begin{equation*}
    p=\frac{1}{N_1}\sum_{i=1}^{N_1}I\{M_{n,\hat{a}}\geq\check{C}_{n,\hat{a}}(\alpha)\}.
\end{equation*}
The previously described procedure is being repeated $n$ times and the average value is taken as the estimated power: 
\begin{equation*}\label{P}
    \tilde{P}=\frac{1}{N}\sum_{i=1}^Np_i.
\end{equation*}

The results are presented in Table \ref{fig: dataDriven20} and \ref{fig: dataDriven50}. The numbers in the parentheses represent the percentage of times that each value of $a$ equaled the estimated optimal one. It is important to note that this bootstrap powers are comparable 
to the maximum achievable power for the tests calculated over a grid of values of the tuning parameter.
 \begin{table}[htbp]
	\centering
	\bigskip
	\caption{Percentage of rejected samples for different value of $a$, $n=20$, $\alpha=0.05$ }
\begin{tabular}{cccccc}
     & 0.5 & 1 & 2 & 5 & $\hat{a}$\\\hline
    $W(1.4)$ & 46 ({\color{red}50}) & 49 (12) & 50 (15) & 48 (23) & 48 \\
    $\Gamma(2)$ & 66 ({\color{red}63}) & 65 (12) & 65 (10) & 63 (15) & 65\\
    $HN$ & 25 ({\color{red}35}) & 28 (14) & 30 (17) & 32 (34) & 29\\
    $U$ & 64 (20) & 72 (9) & 75 (21) & 80 ({\color{red}50}) & 75\\
    $CH(0.5)$ & 19 ({\color{red}37}) & 21 (15) & 22 (17) & 22 (31) & 21\\
    $CH(1)$ & 18 ({\color{red}35}) & 21 (15) & 23 (16) & 23 (34) & 21 \\
    $CH(1.5)$ & 19 ({\color{red}35}) & 20 (11) & 20 (20) & 24 (34) & 21\\
    $LF(2)$ & 35 (33) & 37 (12) & 38 (20) & 41 ({\color{red}35}) & 38\\
    $LF(4)$ & 49 ({\color{red}35}) & 53 (14) & 54 (16) & 54 ({\color{red}35}) & 52\\
    $EW(1.5)$ & 46 (24) & 53 (12) & 56 (20) & 58 ({\color{red}44}) & 54
\end{tabular}
\label{fig: dataDriven20}
\end{table}
\begin{table}[htbp]
	\centering
	\bigskip
	\caption{Percentage of rejected samples for different value of $a$, $n=50$, $\alpha=0.05$ }
\begin{tabular}{cccccc}
     & 0.5 & 1 & 2 & 5 & $\hat{a}$\\\hline
    $W(1.4)$ & 84({\color{red}43}) & 86(19) & 86(16) & 87(22) & 85 \\
    $\Gamma(2)$ & 97({\color{red}68}) & 97(15) & 96(11) & 95(6) & 97\\
    $HN$ & 48(21) & 53(13) & 57(23) & 62({\color{red}43}) & 57\\
    $U$ & 95(31) & 97(12) & 98(20) & 99({\color{red}37}) & 98\\
    $CH(0.5)$ & 34(19) & 37(11) & 41(20) & 44({\color{red}50}) & 41\\
    $CH(1)$ &  33(18) & 37(13) & 41(18) & 46({\color{red}51}) & 41\\
    $CH(1.5)$ & 33(18) & 37(13) & 42(19) & 44({\color{red}50}) &41\\
    $LF(2)$ & 65(20) & 69(12) & 74(24) & 76({\color{red}44}) & 72\\
    $LF(4)$ & 83(25) & 86(16) & 89(20) & 91({\color{red}39}) & 88\\
    $EW(1.5)$ & 81(17) & 87(13) & 89(22) & 93({\color{red}48}) & 89
\end{tabular}
\label{fig: dataDriven50}
\end{table}

\section{Real data examples}
In this section we apply our tests to two real data examples.

The first data set represents inter-occurrence times of the British scheduled data, measured in number of days and listed in the order of their occurrence in time (see \cite{pyke1965}):\\

\noindent 20 106 14 78 94 20 21 136 56 232 89 33 181 424 14\linebreak
 430 155 205 117 253 86 260 213 58 276 263 246 341 1105 50 136.\linebreak 
 
 Applying the algorithm for data-driven tuning parameter we get $\hat{a}=1$.
 The value of the test statistic $M_{31,1}$ is $6.07\times 10^{-4}$, and the corresponding $p$-value is 0.49, so we cannot reject exponentiality in this case.

The second data set represents failure times for right rear breaks on D9G-66A Caterpillar tractors (see \cite{barlow1975}):\\

\noindent
56 83 104 116 244 305 429 452 453 503 552 614 661 673 683 685 753 763 806 \linebreak
834 838 862 897 904 981 1007 1008 1049 1060 1107 1125 1141 1153 1154 1193  \linebreak
1201 1253 1313 1329 1347 1454 1464 1490 1491 1532 1549 1568 1574 1586 1599 \linebreak
1608 1723 1769 1795 1927 1957 2005 2010 2016 2022 2037 2065 2096 2139 2150 \linebreak
2156 2160 2190 2210 2220 2248 2285 2325 2337 2351 2437 2454 2546 2565 2584  \linebreak
2624 2675 2701 2755 2877 2879 2922 2986 3092 3160 3185 3191 3439 3617 3685  \linebreak
3756 3826 3995 4007 4159 4300 4487 5074 5579 5623 6869 7739.\\

Here we get $\hat{a}=0.5$.
 The value of the test statistic $M_{107,0.5}$ is $0.0239$, and the corresponding $p$-value is less than 0.0001, so our test rejects the null exponentiality hypothesis.

\section{Conclusion}
In this  paper we propose new consistent scale-free exponentiality tests based on Puri-Rubin characterization. The proposed tests are shown to be very efficient in Bahadur sense. Moreover, in small sample case, the tests have reasonable to high empirical powers. They also outperform many recent competitor tests in terms of both efficiency and power, which makes them attractive for use in practice.

\section*{Acknowledgement}
This work was supported by the MNTRS, Serbia under  Grant No. 174012 (first  and second  author).

\bibliographystyle{abbrv}
\bibliography{literaturaLaplace}

\begin{thebibliography}{10}

\bibitem{alizadeh2011testing}
H.~Alizadeh~Noughabi and N.~R. Arghami.
\newblock Testing exponentiality based on characterizations of the exponential
  distribution.
\newblock {\em Journal of Statistical Computation and Simulation},
  81(11):1641--1651, 2011.

\bibitem{allison}
J.~Allison and L.~Santana.
\newblock On a data-dependent choice of the tuning parameter appearing in
  certain goodness-of-fit tests.
\newblock {\em Journal of Statistical Computation and Simulation},
  85(16):3276--3288, 2015.

\bibitem{arnold2013exponential}
B.~C. Arnold and J.~A. Villasenor.
\newblock Exponential characterizations motivated by the structure of order
  statistics in samples of size two.
\newblock {\em Statistics \& Probability Letters}, 83(2):596--601, 2013.

\bibitem{bahadur1971}
R.~R. Bahadur.
\newblock {\em Some limit theorems in statistics}.
\newblock SIAM, Philadelphia, 1971.

\bibitem{barlow1975}
R.~E. Barlow and R.~Campo.
\newblock Total time on test processes and applications to failure data
  analysis.
\newblock In {\em Reliability and Fault Tree Analysis}, pages 451--481. SIAM,
  1975.

\bibitem{bozin}
V.~Bo\v{z}in, B.~Milo\v{s}evi\'c, {\relax Ya}.~{\relax Yu}. Nikitin, and
  M.~Obradovi\'c.
\newblock New characterization based symmetry tests.
\newblock {\em Bulletin of the Malaysian Mathematical Sciences Society}, 2018.
\newblock DOI:10.1007/s40840-018-0680-3.

\bibitem{grane2009location}
A.~Gran{\'e} and J.~Fortiana.
\newblock A location-and scale-free goodness-of-fit statistic for the
  exponential distribution based on maximum correlations.
\newblock {\em Statistics}, 43(1):1--12, 2009.

\bibitem{grane2011directional}
A.~Gran{\'e} and J.~Fortiana.
\newblock A directional test of exponentiality based on maximum correlations.
\newblock {\em Metrika}, 73(2):255--274, 2011.

\bibitem{henze1992new}
N.~Henze.
\newblock A new flexible class of omnibus tests for exponentiality.
\newblock {\em Communications in Statistics-Theory and Methods},
  22(1):115--133, 1992.

\bibitem{henze2002goodness}
N.~Henze and S.~G. Meintanis.
\newblock Goodness-of-fit tests based on a new characterization of the
  exponential distribution.
\newblock {\em Communications in Statistics-Theory and Methods},
  31(9):1479--1497, 2002.

\bibitem{henze2002tests}
N.~Henze and S.~G. Meintanis.
\newblock Tests of fit for exponentiality based on the empirical {L}aplace
  transform.
\newblock {\em Statistics: A Journal of Theoretical and Applied Statistics},
  36(2):147--161, 2002.

\bibitem{henze2005}
N.~Henze and S.~G. Meintanis.
\newblock Recent and classical tests for exponentiality: a partial review with
  comparisons.
\newblock {\em Metrika}, 61(1):29--45, 2005.

\bibitem{iverson}
H.~Iverson and R.~Randles.
\newblock The effects on convergence of substituting parameter estimates into
  {U}-statistics and other families of statistics.
\newblock {\em Probability Theory and Related Fields}, 81(3):453--471, 1989.

\bibitem{jevremovic1991note}
V.~Jevremovic.
\newblock A note on mixed exponential distribution with negative weights.
\newblock {\em Statistics \& probability letters}, 11(3):259--265, 1991.

\bibitem{jovanovic}
M.~Jovanovi{\'c}, B.~Milo{\v{s}}evi{\'c}, {\relax Ya}.~{\relax Yu}. Nikitin,
  M.~Obradovi{\'c}, and {\relax K}.~{\relax Yu.}. Volkova.
\newblock Tests of exponentiality based on {A}rnold--{V}illasenor
  characterization and their efficiencies.
\newblock {\em Computational Statistics \& Data Analysis}, 90:100--113, 2015.

\bibitem{klar2003test}
B.~Klar.
\newblock On a test for exponentiality against {L}aplace order dominance.
\newblock {\em Statistics}, 37(6):505--515, 2003.

\bibitem{klar2005tests}
B.~Klar.
\newblock Tests for exponentiality against the {M} and {LM-C}lasses of life
  distributions.
\newblock {\em Test}, 14(2):543--565, 2005.

\bibitem{korolyuk}
V.~S. Korolyuk and Y.~V. Borovskikh.
\newblock {\em Theory of {U}-statistics}.
\newblock Kluwer, Dordrecht, 1994.

\bibitem{meintanis2007testing}
S.~Meintanis, {\relax Ya}.~{\relax Yu}. Nikitin, and A.~Tchirina.
\newblock Testing exponentiality against a class of alternatives which includes
  the {RNBUE} distributions based on the empirical laplace transform.
\newblock {\em Journal of Mathematical Sciences}, 145(2):4871--4879, 2007.

\bibitem{meintanis2008tests}
S.~G. Meintanis.
\newblock Tests for generalized exponential laws based on the empirical
  {M}ellin transform.
\newblock {\em Journal of Statistical Computation and Simulation},
  78(11):1077--1085, 2008.

\bibitem{bojanaMetrika}
B.~Milo{\v{s}}evi{\'c}.
\newblock Asymptotic efficiency of new exponentiality tests based on a
  characterization.
\newblock {\em Metrika}, 79(2):221--236, 2016.

\bibitem{MilosevicObradovicPapers}
B.~Milo{\v{s}}evi{\'c} and M.~Obradovi{\'c}.
\newblock New class of exponentiality tests based on {U}-empirical {L}aplace
  transform.
\newblock {\em Statistical Papers}, 57(4):977--990, 2016.

\bibitem{Publ}
B.~Milo{\v{s}}evi{\'c} and M.~Obradovi{\'c}.
\newblock Some characterization based exponentiality tests and their {B}ahadur
  efficiencies.
\newblock {\em Publications de L'Institut Mathematique}, 100(114):107--117,
  2016.

\bibitem{milovsevic2016some}
B.~Milo{\v{s}}evi\'c and M.~Obradovi\'c.
\newblock Some characterizations of the exponential distribution based on order
  statistics.
\newblock {\em Applicable Analysis and Discrete Mathematics}, 10(2):394--407,
  2016.

\bibitem{nikitin2016efficiency}
Y.~Y. Nikitin and K.~Y. Volkova.
\newblock Efficiency of exponentiality tests based on a special property of
  exponential distribution.
\newblock {\em Mathematical Methods of Statistics}, 25(1):54--66, 2016.

\bibitem{nikitin}
{\relax Ya}.~{\relax Yu}. Nikitin.
\newblock {\em Asymptotic efficiency of nonparametric tests}.
\newblock Cambridge University Press, New York, 1995.

\bibitem{nikitinMetron}
{\relax Ya}.~{\relax Yu}. Nikitin and I.~Peaucelle.
\newblock Efficiency and local optimality of nonparametric tests based on {U}-
  and {V}-statistics.
\newblock {\em Metron}, 62(2):185--200, 2004.

\bibitem{NikVol}
{\relax Ya}.~{\relax Yu}. Nikitin and K.~{\relax Yu.}. Volkova.
\newblock Asymptotic efficiency of exponentiality tests based on order
  statistics characterization.
\newblock {\em Georgian Mathematical Journal}, 17(4):749--763, 2010.

\bibitem{obradovic2014three}
M.~Obradovi{\'c}.
\newblock Three characterizations of exponential distribution involving median
  of sample of size three.
\newblock {\em Journal of Statistical Theory and Applications}, 14(3):257--264,
  2015.

\bibitem{puri1970characterization}
P.~S. Puri and H.~Rubin.
\newblock A characterization based on the absolute difference of two iid random
  variables.
\newblock {\em The Annals of Mathematical Statistics}, 41(6):2113--2122, 1970.

\bibitem{pyke1965}
R.~Pyke.
\newblock Spacings.
\newblock {\em Journal of the Royal Statistical Society. Series B
  (Methodological)}, 27(3):395--449, 1965.

\bibitem{Serfling}
R.~Serfling.
\newblock {\em Approximation theorems of mathematical statistics}, volume 162.
\newblock John Wiley \& Sons, New York, 2009.

\bibitem{volkova2015goodness}
K.~Volkova.
\newblock Goodness-of-fit tests for exponentiality based on
  {Y}anev-{C}hakraborty characterization and their efficiencies.
\newblock {\em Proceedings of the 19th European Young Statisticians Meeting,
  Prague}, pages 156--159, 2015.

\bibitem{yanev2013characterizations}
G.~P. Yanev and S.~Chakraborty.
\newblock Characterizations of exponential distribution based on sample of size
  three.
\newblock {\em Pliska Studia Mathematica Bulgarica}, 22(1):237p--244p, 2013.

\bibitem{Zolotarev}
V.~M. Zolotarev.
\newblock Concerning a certain probability problem.
\newblock {\em Theory of Probability \& Its Applications}, 6(2):201--204, 1961.

\end{thebibliography}

\end{document}